\documentstyle[preprint,aps]{revtex}
\begin{document}

\draft

\title {Universality class of non-Fermi liquid behavior
 in mixed valence systems}

\author{Guang-Ming Zhang}
\address{International Center for Theoretical Physics, P. O. Box 586, 34100
 Trieste, Italy;  \\
and presently Mathematics Department, Imperial College, London SW7 2BZ, U.K.}
\author{Zhao-Bin Su and Lu Yu}
\address{International Center for Theoretical Physics, P. O. Box 586, 34100
 Trieste, Italy; \\ and Institute of Theoretical Physics, Academia
Sinica, Beijing 100080, China.}

\maketitle

\begin{abstract}
A generalized Anderson single-impurity model with off-site Coulomb interactions
is derived from the extended three-band Hubbard model, originally
proposed to describe the physics of the  copper-oxides. Using the abelian
bosonization technique and canonical transformations, an effective Hamiltonian
is derived in the strong coupling limit, which is essentially analogous to
the Toulouse limit of the ordinary
Kondo problem. In this limit, the effective Hamiltonian can be exactly solved,
with a mixed valence quantum critical point separating two different
Fermi liquid phases, {\it i.e.} the Kondo phase and the empty orbital phase. In
the  mixed valence quantum critical regime, the local moment is only partially
quenched and X-ray edge singularities are generated. Around the quantum
critical point, a new type of non-Fermi liquid behavior is predicted with
an extra specific heat $C_{imp}\sim T^{1/4}$ and a singular
spin-susceptibility $\chi_{imp}\sim T^{-3/4}$. At the same time,
the effective Hamiltonian under single occupancy is transformed into a
resonant-level model, from which the correct Kondo physical properties
(specific heat, spin susceptibility, and  an enhanced Wilson ratio) are easily
rederived. Finally, a brief discussion is
given to relate these theoretical results to observations in $UPd_xCu_{5-x}$
($x=1,1.5$) alloys, which show single-impurity critical behavior consistent
with
our predictions.
\end{abstract}


\newpage
\section{Introduction}

Correlated electron systems ranging from heavy-fermions  to
high-temperature superconductors exhibit a fascinating variety of
anomalous behavior. An
issue of considerable current debate is to what extent these systems in the
normal state can be described as Fermi liquids (FL) where the low-energy
excitations have a one-to-one correspondence to those of a noninteracting Fermi
gas with the well-known behavior of a specific heat $C=\gamma T$, a
temperature independent Pauli
susceptibility $\chi$, and a low temperature electrical
resistivity $\rho\approx AT^2$. Most previous investigations on heavy-fermion
systems, in particular, on the mixed valence problem are carried out within the
framework of the FL theory. In the meantime a phenomenological marginal
Fermi liquid (MFL) theory\cite{vlsar},
proposed for high-temperature superconductors to explain certain
features of the observed diversified anomalies, has brought about a lot of new
developments in this active field of research.

Recently,  a lot of strong  evidence of non-FL behavior has been reported for
several heavy fermion systems \cite{rev}, including  $U_xY_{1-x}Pd_3$
\cite{maple} and $UPd_xCu_{5-x}$ alloys \cite{andraka,aronson,tsve}.
In some of these systems, the non-FL behavior may be due to a $T=0$
quantum phase transition of cooperative origin\cite{rev}, while in the others
the singular behavior of thermodynamical and transport properties
is consistent with a single-impurity critical scaling\cite{tsve}.
Since no reliable methods have been devised so far to study the microscopic
mechanism for the non-FL behavior in any lattice models, it is thus expected
that the investigations of the related impurity models would shed some light on
the basic physics of the lattice models, in the same sense as the original
Anderson
single-impurity model vs the single band Hubbard model (or Anderson lattice
model). Various impurity models have been studied to explore the possible
non-FL behavior
\cite{nb,adtwss,al,eksg,giamarchi,zsy1,perakis,zy,sire,zsy2,si,gabi}.
 In particular, the two-channel Kondo system, intensively studied using
different methods \cite{nb,adtwss,al,eksg} displays a kind of MFL behavior.
However, it is not clear at all how to directly relate such a
two-channel Kondo model to the real correlated f-electron heavy-fermion systems
and high-temperature superconductors. In fact, a long standing puzzle of
the mixed valence problem, namely to understand the complexity of phenomena
in realistic systems in terms of fundamental physical concepts \cite{and87},
is  still not fully resolved. However, it seems to us that the
key issue is whether the FL picture  is correct for the mixed valence state.

The mixed valence phenomena \cite{lrp} usually occur in rare-earth compounds in
which the proximity of the 4f-level to the Fermi surface leads to
 substantial
charge (valence) fluctuations and  instabilities  of the magnetic moment.
 Basically, the mixed
valence state can be thought of as a mixture of two bonding states
$4f^n(5d6s)^m$
and $4f^{n-1}(5d6s)^{m+1}$, which are nearly degenerate from the quantum
mechanical point of view. This situation must be understood in terms of the
hybridization of the  above two configurations so that the hybridized level
is only partially occupied in the low-energy excitations. Although such a
situation bears a strong resemblance to the crossover of the single-impurity
problem from free moment behavior at high temperature to a strong-coupling FL
regime in the Kondo ground state. However, that crossover does not coincide
with significant  valence change in most mixed valence systems \cite{lrp}.
It was argued that the
Friedel sum rule (in the sense of local charge neutrality) is important
for the mixed valence problem and it should be satisfied for both valence
states.
In this sense  the usual Anderson model is not complete (because
the Friedel sum rule cannot be satisfied
for two different
valence states by varying only one parameter (the $f-$ level)), so  one
should  extend it to
include the screening mechanism from the $5d6s$ electrons to describe the $4f$
charge fluctuations \cite{haldane,and87}.

Recently, the interest to this problem has been revived in connection with high
$T_c$ superconductors. Varma and others \cite{perakis,zy,sire,zsy2}
have considered a
generalized Anderson model with finite-range Coulomb interactions,
equivalent to a
single-site version of the extended three-band Hubbard model proposed
originally to
describe the physics of copper-oxide superconductors. The numerical evidence
for a mixed-valence quantum critical point came from a Wilson renormalization
group study\cite{perakis}, while a strong coupling limit was argued and
considered for this critical point later\cite{sire}. Inspired by these
studies, two of us\cite{zy} have shown that using the renormalization-group
 analysis a phase diagram   with a mixed valence quantum critical
regime separating the Kondo phase and the empty orbital phase,  can be deduced.
It has  also been shown\cite{zy} that the strong coupling limit can be
derived straightforwardly from the original Hamiltonian using the canonical
transformations and an  exact solution to this model
can be found in this  limit.
However, the physics behind the mixed valence state has not been fully
revealed. The two basic important questions are: How to derive a complete
effective model Hamiltonian in the strong coupling limit and what are the
generic physical properties of the mixed valence state?

 In the present paper, we address these unsolved problems in the infinite-$U$
generalized Anderson single-impurity model with finite-range Coulomb
interactions. Through two successive canonical transformations in the bosonic
representation, we can derive an effective Hamiltonian in
a particular strong coupling limit, which is analogous to the Toulouse limit
of the ordinary Kondo problem\cite{tou}. In this limit, the effective
Hamiltonian can be exactly solved, with a mixed valence quantum critical
point separating two different Fermi
liquid phases, {\it i.e.} the Kondo phase and the empty orbital phase. In this
mixed valence quantum critical regime, the local moment is only partially
quenched and X-ray edge singularities are generated. Around this
critical point, a new
type of non-Fermi liquid behavior is predicted with an extra specific heat
$C_{imp}\sim T^{1/4}$ and a singular spin-susceptibility
$\chi_{imp}\sim T^{-3/4}$. In the meantime, the effective
Hamiltonian under
 single occupancy is transformed into a resonant-level model \cite{wfs},
from which the
correct Kondo physical properties (specific heat, spin susceptibility,
and enhanced Wilson ratio) can be easily rederived. Finally, a brief
discussion is
given to relate our results with the experimental observations in
$UPd_xCu_{5-x}$ ($x=1,1.5$) alloys.

The arrangement of the paper is as follows: We first derive the model
Hamiltonian
from the extended three-band Hubbard model  in Section II.
Then, in Section III a non-perturbative approach for this single
impurity model is
 formulated and a complete effective Hamiltonian is
presented. In Section IV, the exact calculations of the physical properties
in the strong-coupling limit are presented. Finally,
discussions on experiments
are given in Section V, while concluding remarks
are made in Section VI.

\section{Derivation of the local impurity model from the
extended three-band Hubbard model}

The extended three-band Hubbard model is based on the observation that
the chemistry of the  copper-oxides is unique in that
charge fluctuations between the cations and the anions controlled by
their mutual interactions occur at low-energies \cite{ev}. Such dynamical
effects
were expected to qualitatively change the low-energy response,
compared with the single band Hubbard model and its large Hubbard
$U$ limit, the $t-J$ model.  The model is composed of $d$ hole states on
the copper
sites (one hole per copper in the undoped material) in $d_{x^2-y^2}$
orbitals which are strongly hybridized with $p$ $\sigma$-orbitals on the
neighboring oxygen sites. As the undoped material is a magnetic
insulator, the holes on the copper sites must be localized by a reasonably
strong on-site Coulomb interaction. Also included are the finite-range
Coulomb repulsions between oxygen electrons on nearest-neighbor sites
of the localized copper electrons. The corresponding Hamiltonian is
given by:
\begin{eqnarray}
&& H=\sum_{i,\sigma}\epsilon_d
d^{\dag}_{i,\sigma}d_{i,\sigma}
+ \sum_{\vec{k},\sigma}(\epsilon_{\vec{k}}-\mu)
p^{\dag}_{\vec{k},\sigma}p_{\vec{k},\sigma} + U
\sum_{i}d^{\dag}_{i,\uparrow}d_{i,\uparrow}
d^{\dag}_{i,\downarrow}d_{i,\downarrow}
\nonumber \\ && \hspace{.5cm}
+\frac{1}{\sqrt{N}}\sum_{i,\vec{k},\sigma}(t_{i,\vec{k}}
p^{\dag}_{\vec{k},\sigma}d_{i,\sigma}+h.c.) +
\frac{1}{2N}\sum_{i,\vec{k},\vec{k'},\sigma}
V_{i,\vec{k},\vec{k}'}
(p^{\dag}_{\vec{k},\sigma}p_{\vec{k}',\sigma}-
p_{\vec{k}',\sigma}p^{\dag}_{\vec{k},\sigma})
(\sum_{\sigma'}d^{\dag}_{i,\sigma'}d_{i,\sigma'} - \frac{1}{2}),
\label{threeband}\end{eqnarray}
where $p_{\vec{k},\sigma}$ is the Bloch representation of the
oxygen orbitals, while $V_{i,\vec{k},\vec{k}'}$ is the
Fourier transform of the Coulomb interactions of the $i-$ site
copper with neighboring oxygens, and $N$ is the total number of
copper-oxygen cells. The rest of notation has obvious meaning.
As pointed out in the Introduction, the above lattice model has so far
proven intractable to give us reliable information on the low-energy
physics of the system. However, one can regard strong coupling
in the lattice models as a low-dimensional critical phenomenon involving
long-time fluctuations at each localized site, but no critical
fluctuations in space. The long-time fluctuations at each site are
independent because of the local conservation laws
when coherent effects and inter-impurity interactions are weak.
Therefore, one can use a single-impurity model to
describe the copper-oxides or heavy-fermion metals under the physical
hypothesis that the essential new physics is local in real space, on the
scale of a lattice constant, so all the interesting effects are in the time
or frequency domain. If the singularities found in the impurity
problem do not depend on any
special symmetries lost when going from the single impurity to the lattice,
they are likely to be
relevant to the behavior of the full lattice problem as well. Consider now
 the single-site
(copper site) version of Eq.\ (\ref{threeband}) in the following form:
\begin{eqnarray}
&& H=
\sum_{\sigma}\epsilon_d d^{\dag}_{\sigma}d_{\sigma} +
\sum_{\vec{k},\sigma}(\epsilon_{\vec{k}}-\mu)
p^{\dag}_{\vec{k},\sigma}p_{\vec{k},\sigma} + U
d^{\dag}_{\uparrow}d_{\uparrow}d^{\dag}_{\downarrow}d_{\downarrow}
 \nonumber \\ && \hspace{.5cm}
+ \frac{1}{\sqrt{N}}\sum_{\vec{k},\sigma}(t_{\vec{k}}
p^{\dag}_{\vec{k},\sigma}d_{\sigma}+h.c.) +
\frac{1}{2N}\sum_{\vec{k},\vec{k}',\sigma}V_{\vec{k},\vec{k'}}
(p^{\dag}_{\vec{k},\sigma}p_{\vec{k}',\sigma}-
p_{\vec{k}',\sigma}p^{\dag}_{\vec{k},\sigma})
(\sum_{\sigma'}d^{\dag}_{\sigma'}d_{\sigma'} -\frac{1}{2}) ,
\label{singlesite} \end{eqnarray}
where we set the single impurity at the origin of the coordinates
and, as usual, assume the
$d_{x^2-y^2}$ orbitals are non-degenerate so that the quantum
numbers $(l_d,m_d)$ describing the orbital angular momentum, are fixed. It
should be pointed out that this single-impurity model is still a
three-dimensional system with two different angular momentum orbitals of the
$d$-electrons and $p$-electrons. In the following, we have to reduce this
three dimensional model to a one-dimensional model by projecting the
$p$ $\sigma$-orbitals onto the $d_{x^2-y^2}$ orbitals.

First of all, we transform the plane wave representation of the
$p$-electrons to the spherical representation in view of the assumed
 spherical symmetry about the impurity site.
\begin{equation}
 p_{\vec{k},\sigma}=\sum_{l,m}<\hat{k}\mid l,m>C_{k,\sigma,l,m},
\nonumber \end{equation}
where we have used the definition $\vec{k}=(k,\hat{k})$ and the quantum
number $m$ can take values  $-l,-l+1, ...,+l$. If we project the $p$
$\sigma$-orbitals onto the $d$ orbitals, the angular quantum number $l$
is required to take only one value, that is the angular quantum number
 of the $d$
orbital $l=l_d$. In what follows, this quantum number is dropped,
and the quantum number  $m$ is regarded as a
"channel" index of the $p$ orbital electrons. Since the dispersion
relation of the $p$-electrons also has spherical symmetry, the kinetic
energy term in Eq.\ (\ref{singlesite})
$\sum_{\vec{k},\sigma}(\epsilon_{\vec{k}}-\mu)
p^{\dag}_{\vec{k},\sigma}p_{\vec{k},\sigma}$ is transformed into
$\sum_{k,\sigma,m}(\epsilon_{k}-\mu)C^{\dag}_{k,\sigma,m}C_{k,\sigma,m}$,
which implies there are several conduction electron channels
 in the reduced model.
At the same time, the hybridization term between the $d$- and $p$-electrons
$\frac{1}{\sqrt{N}}\sum_{\vec{k},\sigma}(t_{\vec{k}}
p^{\dag}_{\vec{k},\sigma}d_{\sigma}+h.c.)$ is changed into
$\frac{1}{\sqrt{N}}\sum_{k,\sigma}
(t_k C^{\dag}_{k,\sigma,m_d}d_{\sigma} +h.c.)$. As required by symmetry,
the localized $d$ orbitals can hybridize only with $p$ orbitals with the same
angular quantum numbers. Due to the non-degeneracy of the
$d$ orbitals, there is only one channel with $m=m_d$ of $p$-electrons
hybridizing with the local $d$ orbitals, which in the following is called
hybridizing channel and the others are called screening channels.

Second, the finite range Coulomb interaction  between the localized
orbitals and the delocalized orbitals, or the  X-ray edge
(XRE) like potential scattering,
is considered. Due to the spherical symmetry about the impurity
site, the scattering potential $V_{\vec{k},\vec{k'}}$  depends only on
the angle between $\hat{k}$ and $\hat{k'}$, so it can be expanded as
$$\sum_{l,m}V_{k,k',l,m}<\hat{k}\mid l,m><l,m\mid\hat{k'}>,$$
where the only remaining term in the summation
over the angular momentum is $l=l_d$.
Finally, putting everything together, the complete form of the reduced
one-dimensional single-impurity
Hamiltonian is obtained as:
\begin{eqnarray}
&&H= \sum_{k,\sigma,m}(\epsilon_{k}-\mu)C^{\dag}_{k,\sigma,m}C_{k,\sigma,m}
+ \sum_{\sigma}\epsilon_d d^{\dag}_{\sigma}d_{\sigma} +
U d^{\dag}_{\uparrow}d_{\uparrow}d^{\dag}_{\downarrow}d_{\downarrow}
+\frac{t}{\sqrt{N}}\sum_{k,\sigma}(C^{\dag}_{k,\sigma,m_d}d_{\sigma}+h.c.)
\nonumber \\ && \hspace{.5cm}
+\frac{1}{2N}\sum_{k,k',\sigma}\sum_{\{m=-l_d,...+l_d\}} V_{m}
(C^{\dag}_{k,\sigma,m}C_{k',\sigma,m}- C_{k',\sigma,m}C^{\dag}_{k,\sigma,m})
 (\sum_{\sigma'}d^{\dag}_{\sigma'}d_{\sigma'} - \frac{1}{2}),
\label{effect} \end{eqnarray}
where, without loss of any generality, the hybridization strength $t_k$
 and the Coulomb interaction parameters $V_{k,k',m}$ have been further
assumed to be momentum independent,  as in the usual treatments of the Anderson
single-impurity model. This  is the so-called local copper-oxide model or
generalized Anderson single-impurity model with finite-range Coulomb
interactions used to describe the local version of non-FL properties of
the copper-oxide compounds\cite{perakis,zy,sire}.
 Here we have presented an explicit derivation.
Although the derivation itself is given in terms of copper-oxygen
orbitals, valid only for cuprates, the charge fluctuation physics,
incorporated here via including the $d-$orbitals and finite range Coulomb
interactions between $d$ and $p$ electrons,
is also materialized in other mixed valence compounds, especially
in  heavy fermion systems. Therefore, this model can be applied
to these systems as well. It is worth mentioning that the
infinite dimensional technique has also been  used to study the above
extended three-band Hubbard model \cite{si}, and  a
single-channel generalized Anderson single-impurity model with
self-consistent condition has been derived. Since the infinite-dimensional
approximation
overemphasizes the dimensional symmetries, it, probably, will
miss some other important symmetry properties in the
 considered correlated electron
systems, and the most important mixed valence physics we are concerned
with (the
screening effects) is not included in that generalized Anderson
single-impurity model.

\section{Bosonization for the generalized Anderson single-impurity model}

Usually, the interesting low-energy physics contained in the above
single-impurity model involves only  two configurations: $n_d=0$ and
$n_d=1$, and we can remove the $n_d=2$ configuration by letting
$U\rightarrow \infty$. According to the numerical RG analysis for the usual
Anderson impurity model, apart from the particle-hole symmetric case,
 reducing to the Kondo model,
particular attention should be paid to the asymmetric case \cite{krishna80}.
In the infinite-U limit, the model Hamiltonian is defined by $H=H_h+H_s$,
where
\begin{eqnarray}
&& H_h=\sum_{k,\sigma}\epsilon_{k}C^{\dag}_{k,\sigma,0}C_{k,\sigma,0}
+\epsilon_{d}n_d+ \frac{h}{2}\sum_{\sigma}\sigma n_{d,\sigma}
+t\sum_{\sigma}(C^{\dag}_{\sigma,0}d_{\sigma}+h.c.)
+ \frac{h}{2}\sum_{k,\sigma} \sigma C^{\dag}_{k,\sigma,0}C_{k,\sigma,0}
 \nonumber \\ && \hspace{1cm}
+ V \sum_{\sigma}(C^{\dag}_{\sigma,0}C_{\sigma,0}-\frac{1}{2})(n_{d}-1)
+\frac{J}{4}\sum_{\sigma,\sigma'}
C^{\dag}_{\sigma,0}\vec{\sigma}C_{\sigma',0}
\sum_{\mu,\nu}d^{\dag}_{\mu}\vec{\sigma}d_{\nu}
\nonumber \\ &&
 H_s=\sum_{k,\sigma,m>0}\epsilon_{k}C^{\dag}_{k,\sigma,m}C_{k,\sigma,m}
+\sum_{\sigma,m>0}V_m (C^{\dag}_{\sigma,m}C_{\sigma,m}-\frac{1}{2})
(n_d-\frac{1}{2})+ \frac{h}{2} \sum_{k,\sigma,m>0}\sigma
C^{\dag}_{k,\sigma,m}C_{k,\sigma,m},
\end{eqnarray}
where $C_{\sigma,m}=\frac{1}{\sqrt{N}}\sum_k C_{k,\sigma,m}$ is the
conduction electron operator at the origin of the coordinates,
$n_d=\sum_{\sigma}n_{d,\sigma}$, $N$ is the number of the lattice
sites, and we have separated the Hamiltonian into the hybridizing ($m=0$) and
screening parts ($m{\not=}0$). The chemical potential of the conduction
electrons has been chosen to be zero and the energy level of the local impurity
$\epsilon_{d}$ is also assumed to be very close to it
 because of the character of the mixed valence
state. We include a uniform external magnetic field $h$ to
calculate the total spin susceptibility enhancement. The infinite $U$ limit
keeps the low-energy physics of the
model, but we should work under a local constraint $n_d\leq 1$, which
is really crucial for determining the physical behavior of the
low-energy excitations. Equally, we can rewrite $H_h$ in another form, which is
useful for the following analysis.
\begin{eqnarray}
&& H_h=\sum_{k,\sigma}\epsilon_{k}C^{\dag}_{k,\sigma,0}C_{k,\sigma,0}
+\epsilon_{d}n_d+ \frac{h}{2}\sum_{\sigma}\sigma n_{d,\sigma}
+t\sum_{\sigma}(C^{\dag}_{\sigma,0}d_{\sigma}+h.c.)
+ \frac{h}{2}\sum_{k,\sigma} \sigma C^{\dag}_{k,\sigma,0}C_{k,\sigma,0}
\nonumber \\ &&
+V_0\sum_{\sigma}(C^{\dag}_{\sigma,0}C_{\sigma,0}-\frac{1}{2})
(n_{d,\sigma}-\frac{1}{2})
+V'_0\sum_{\sigma}(C^{\dag}_{\sigma,0}C_{\sigma,0}-\frac{1}{2})
(n_{d,{\bar \sigma}}-\frac{1}{2}) + V_{\perp}\sum_{\sigma}
C^{\dag}_{\sigma,0}C_{\bar{\sigma},0}d^{\dag}_{\bar{\sigma}}d_{\sigma},
\end{eqnarray}
where $V_0=(V+\frac{J_z}{4})$ is the parallel spin scattering potential,
 $V'_{0}=(V-\frac{J_z}{4})$ is the opposite spin scattering potential,
and $V_{\perp}=\frac{J_{\perp}}{2}$ is the spin-flip scattering potential.

We first bosonize the screening part. Note that it has only one Fermi
point for each channel and the dispersion is linearized
$\epsilon_k=(k-k_F)/\rho$ with a cutoff $k_D$ and $\rho=(hv_F)^{-1}$,
so  we can
define the bosonic operators as\cite{ss},
$$ b_{k,\sigma,m}= \frac{1}{\sqrt{kN}}\sum_{q=0}^{k_D-k}
C^{\dag}_{q,\sigma,m}C_{q+k,\sigma,m}, \hspace{.5cm}
  b^{\dag}_{k,\sigma,m}= \frac{1}{\sqrt{kN}}\sum_{q=k}^{k_D}
C^{\dag}_{q,\sigma,m}C_{q-k,\sigma,m}, \hspace{.5cm} k> 0, $$
which obey the standard commutation relations. Since the spin degrees of
freedom of the screening channel electrons are trivially involved, they
can be separated from $H_s$, which then becomes
\begin{eqnarray}
&& H_s^{b}=\sum_{k,m>0}\frac{k}{\rho}(a^{\dag}_{k,m}a_{k,m}+
 e^{\dag}_{k,m}e_{k,m})+ \sum_{k,m>0}{V_m}\sqrt{\frac{2k}{N}}
(a^{\dag}_{k,m}+a_{k,m})(n_d-\frac{1}{2}) \nonumber \\ && \hspace{1cm}
+ \frac{\sqrt{2}h}{4\pi}\sum_{k>0,m}\sqrt{\frac{k}{N}}
\int dx [(e_{k,m}e^{ikx}+ e_{k,m}^{\dag}e^{-ikx})],
\end{eqnarray}
where
$a_{k,m}=\frac{1}{\sqrt{2}}(b_{k,\uparrow,m}+b_{k,\downarrow,m})$ and
 $e_{k,m}=\frac{1}{\sqrt{2}}(b_{k,\uparrow,m}-b_{k,\downarrow,m})$
are the charge- and spin-density operators, respectively. Moreover, we
can assume $V_m=V_s$ for all $m>0$ without loss of generality, so the
channel index can be dropped. Thus, the screening part is
simplified as a single spinless channel:
\begin{equation}
H_s^{b}=
\sum_{k>0}\frac{k}{\rho}a^{\dag}_{k}a_{k}
+\sum_{k>0}{\tilde{V}_s}\sqrt{\frac{k}{N}}(a^{\dag}_{k}+
a_{k})(n_d-\frac{1}{2}),
\end{equation}
where we have dropped the spin-density operators because they are not
involved in the interactions with the local impurity, while the
charge-density part is kept:
$a_k=\frac{1}{\sqrt{N_s}}\sum_{m>0}a_{k,m}$,
$\tilde{V}_s\equiv\sqrt{2N_s}V_s$ and $N_s$ is the number of the
screening channels. Using the inverse bosonization, we can transform
bosons back to fermions:
\begin{equation}
H_s^{f}=\sum_{k}\epsilon_{k}s^{\dag}_{k}s_{k}
+\frac{\tilde{V}_s}{2N}\sum_{k,k'}(s^{\dag}_{k}s_{k'}-s_{k'}s^{\dag}_{k})
(n_d-\frac{1}{2}).
\end{equation}
Interestingly enough, the  strong electron-phonon interaction accompanying a
valence change on a localized impurity bears the same form as $H_s^b$. So
in the
abelian representation, $H_s$ can be thought of including the effects due to
the large  cell volume change ($15$ percent) accompanying the charge
fluctuations,
which is also expected to significantly alter the dynamics in the mixed valence
problem. Therefore, the physics we will discuss in the strong-coupling limit
to some extent reflects some essential
aspects of the real mixed valence systems.

The same bosonization procedures can be used for the hybridizing
electrons as well, and the boson operators are defined as before.
Restricted to the low-lying excitations, the hybridizing
Hamiltonian is expressed as:
\begin{eqnarray}
&&H_h^b=
\sum_{k>0,\sigma}\frac{k}{\rho}b^{\dag}_{k,\sigma}b_{k,\sigma}
+\epsilon_{d}n_d+\frac{h}{2}(n_{d,\uparrow}-n_{d,\downarrow})
+t\sum_{\sigma}(C^{\dag}_{\sigma}d_{\sigma}+h.c.)
+V_{\perp}\sum_{\sigma}C^{\dag}_{\sigma}C_{\bar{\sigma}}d^{\dag}_{\bar\sigma}
 d_{\sigma}
\nonumber \\ && \hspace{1cm}
+V_0\sum_{k>0,\sigma}\sqrt{\frac{k}{N}}(b^{\dag}_{k,\sigma}+b_{k,\sigma})
(n_{d,\sigma}-\frac{1}{2})
+{V'}_0\sum_{k>0,\sigma}\sqrt{\frac{k}{N}}(b^{\dag}_{k,\sigma}+b_{k,\sigma})
(n_{d,{\bar \sigma}}-\frac{1}{2})+ \nonumber \\ && \hspace{1cm}
+\frac{h}{4\pi}\sum_{k>0}\sqrt{\frac{k}{N}}\int dx
[(b^{\dag}_{k,\uparrow}e^{ikx} + b_{k,\uparrow}e^{-ikx})-
 (b^{\dag}_{k,\downarrow}e^{ikx} + b_{k,\downarrow}e^{-ikx})].
\end{eqnarray}
Here $C_{\sigma}(x)=\sqrt{k_D}exp\{\sum_{k>0}\frac{1}{\sqrt{kN}}
(b_{k,\sigma}e^{ikx}-b^{\dag}_{k,\sigma}e^{-ikx})\}$ is the fermion expression
in terms of the bosons.

In order to derive an effective Hamiltonian in the Toulouse
limit, we perform the following two canonical transformations
 $$ U=exp\{\sum_{k>0}\frac{1}{\sqrt{kN}}(a_k-a_k^{\dag})(n_d-\frac{1}{2})\},
 \hspace{.5cm}
 S= exp\{\sum_{\{k>0,\sigma\}}\frac{1}{\sqrt{kN}}
(b_{k,\sigma} - b_{k,\sigma}^{\dag}) (n_{d,\sigma} - \frac{1}{2})\}.$$
As a matter of fact, the first canonical transformation transfers the
singularities of the screening channel into the hybridizing channel via
the local impurity. The second canonical transformation not only makes
the phase shifts due to hybridization and part of the parallel-spin
XRE
scattering compensate each other, but also simplifies the spin-flip
scattering in
the hybridizing channel. This way the hybridizing electrons become free for
a special strong coupling limit, which shares the same physical meaning as the
ordinary Toulouse limit of the usual single-impurity single-channel Kondo
model. The transformed model Hamiltonian is given by:
\begin{eqnarray}
&& \tilde{H}=
\sum_{k>0,\sigma}\frac{k}{\rho}b^{\dag}_{k,\sigma}b_{k,\sigma}
+(\epsilon_d-V'_0k_D)n_d+t\sum_{\sigma}(d_{\sigma}s_0+h.c.)+
V_{\perp}k_D\sum_{\sigma}d^{\dag}_{\bar{\sigma}}d_{\sigma}
+\sum_{k>0}\frac{k}{\rho}a^{\dag}_{k}a_{k}
\nonumber \\ && \hspace{.5cm}
+(V_0-\frac{1}{\rho})\sum_{k>0,\sigma}\sqrt{\frac{k}{N}}
(b^{\dag}_{k,\sigma}+b_{k,\sigma}) (n_{d,\sigma}-\frac{1}{2})
+{V'}_0\sum_{k>0,\sigma}\sqrt{\frac{k}{N}}(b^{\dag}_{k,\sigma}+b_{k,\sigma})
(n_{d,{\bar \sigma}}-\frac{1}{2})+
\nonumber \\ && \hspace{.5cm}
+ \frac{h}{4\pi}\sum_{k>0}\sqrt{\frac{k}{N}}\int dx
[(b_{k,\uparrow} - b_{k,\downarrow})e^{ikx}+
 (b^{\dag}_{k,\uparrow}e^{ikx} - b^{\dag}_{k,\downarrow})e^{-ikx}]
\nonumber \\ && \hspace{.5cm}
+(\tilde{V_s}-\frac{1}{\rho})\sum_{k>0}\sqrt{\frac{k}{N}}
(a^{\dag}_{k}+a_{k})(n_d-\frac{1}{2}),
\end{eqnarray}
Note that the $\frac{h}{2}(n_{d,\uparrow}-n_{d,\downarrow})$ term in
$H_h$ has been exactly canceled by the canonical transformation $S$,
which is similar to the Emery-Kivelson approach for the two-channel Kondo
system \cite{eksg}. A fermionic form is obtained by the inverse bosonization
${\tilde{H}_{eff}=H_T+\delta H}$, where
\begin{eqnarray}
&& H_T =
\sum_{k,\sigma} ( \epsilon_k + \frac{h}{2}\sigma)
C^{\dag}_{k,\sigma}C_{k,\sigma}
+ ( \epsilon_{d} - V'_0 k_D) n_d +
V_{\perp} k_D \sum_{\sigma} d^{\dag}_{\bar{\sigma}}d_{\sigma}
\nonumber \\ && \hspace{1cm}
+ t\sum_{\sigma} (d_{\sigma} s_0 + h.c.)
+\sum_{k} \epsilon_{k} s^{\dag}_{k}s_{k},
\nonumber  \\ &&
\delta H= (V_0-\frac{1}{\rho})
\sum_{\sigma} (C^{\dag}_{\sigma}C_{\sigma}-\frac{1}{2})
(n_{d,\sigma}-\frac{1}{2})
+ V'_0 \sum_{\sigma} (C^{\dag}_{\sigma}C_{\sigma}-\frac{1}{2})
 (n_{d, \bar{\sigma}}-\frac{1}{2})
\nonumber \\ && \hspace{1cm}
+(\tilde{V}_s-\frac{1}{\rho})(s^{\dag}_{0}s_{0}- \frac{1}{2})
(n_d-\frac{1}{2}).
\end{eqnarray}
 This effective Hamiltonian, when
$V_0=\tilde{V}_s=\frac{1}{\rho}$ and $V'_0=0$, reduces to $H_T$, which is the
same as the strong coupling Hamiltonian argued by the renormalization-group
analysis \cite{zy}. It is obvious that $H_T$ is  analogous to the
Toulouse limit
of the ordinary Kondo problem. It should be emphasized that
the full effective model Hamiltonian in the strong coupling limit
is derived here exactly, from which the corresponding physical properties in
 different phases can be calculated explicitly.

\section{Physical properties of the model Hamiltonian in the Toulouse
 strong coupling limit}

\subsection{Exact solution of the Toulouse limit Hamiltonian}

Now, let us first study the strong-coupling limit Hamiltonian $H_T$ with
$V_0=\tilde{V}_s=\frac{1}{\rho}$ and $V'_0=0$,
where only part of degrees of freedom of the local impurity
$\alpha\equiv\frac{1}{\sqrt{2}}(d_{\uparrow}+d_{\downarrow})$ is
coupled to the conduction electrons, while the remaining part
$\beta\equiv\frac{1}{\sqrt{2}}(d_{\uparrow}-d_{\downarrow})$ is decoupled
except for the constraint \cite{zy},\cite{sire}. Thus, the effective
Hamiltonian is:
\begin{eqnarray}
&& H_T= \sum_{k,\sigma} ( \epsilon_k + \frac{h}{2}\sigma)
 C^{\dag}_{k,\sigma}C_{k,\sigma}+\sum_{k}\epsilon_k s^{\dag}_{k}s_{k}+
     (\epsilon_d -V_{\perp}k_D)n_{\beta}
\nonumber \\ && \hspace{1cm}
+(\epsilon_{d}+V_{\perp}k_D)n_{\alpha}+\sqrt{2}t(s_0^{\dag}\alpha^{\dag}+h.c.),
\end{eqnarray}
where the local constraint $n_{\alpha}+n_{\beta}\le 1$ will play a
crucial role in determining the behavior of the system, and thus we must
handle it exactly. In impurity zero-occupancy, the hybridizing and
screening electrons are completely free with a zero phase shift due to
the Friedel sum rule. In impurity single-occupancy, the localized impurity
disappears, but the total energy of the system has a  shift
$(\epsilon_d - V_{\perp}k_D)$, and according to the Friedel sum rule the
hybridizing electrons and the screening electrons have a unitarity phase
shift $\frac{\pi}{2}$ per spin. This corresponds to the behavior of the
Kondo model exactly at its FL fixed point. If we want to calculate the
physical properties, the leading irrelevant interactions should be
involved, equivalent to driving the system away from its fixed point
\cite{wilson}.

When the impurity occupancy fluctuates between zero and one,
 it  requires that $\epsilon_d\approx -V_{\perp}k_D$,
corresponding to the mixed valence quantum critical point found in
Ref.{\cite{perakis}}. The  crucial point is that $H_T$ conserves
$n_{\beta}$ {\cite{zy,sire}}, which makes the following calculations available.
The strong coupling limit Hamiltonian in the subspaces
of $n_{\beta}=0,1$ are the following, without any constraints
\begin{equation}
 H_{n_{\beta}=0} = \sum_{k,\sigma} ( \epsilon_k + \frac{h}{2}\sigma)
 C^{\dag}_{k,\sigma}C_{k,\sigma}+\sum_{k}\epsilon_k s^{\dag}_{k}s_{k}+
(\epsilon_{d}+V_{\perp}k_D)n_{\alpha}+\sqrt{2}t (s_0^{\dag}\alpha^{\dag}+h.c.),
\end{equation}
\begin{equation}
 H_{n_{\beta}=1}= \sum_{k,\sigma} (\epsilon_k + \frac{h}{2}\sigma)
 C^{\dag}_{k,\sigma}C_{k,\sigma}+ \sum_{k}\epsilon_k s^{\dag}_{k}s_{k}
  +(\epsilon_d -V_{\perp}k_D).
\end{equation}
Thus, we can carry out our calculations in the subspaces of $n_{\beta}$.
Any eigenstates of $H_T$ should be decomposed of the known eigenstates
in the subspaces $n_{\beta}=0,1$:
\begin{equation}
\mid\Psi>=\mid\phi>_{ n_{\beta}=0}+\mid\phi>_{n_{\beta}=1}.
\end{equation}
In the $n_{\beta}=0$ subspace, the constraint is
satisfied for any $\alpha$, so the quadratic Hamiltonian
can be exactly solved, and a useful expectation associated with $\alpha$
is derived as:
\begin{equation}
 <T_{\tau}\alpha(\tau)\alpha^{\dag}(0)>_{n_{\beta}=0}=-T\sum_n
\frac{e^{i\omega_n\tau}} {i\omega_n-\epsilon_{\alpha}+i\Gamma sign(\omega_n)}
\approx \frac{1}{\pi\Gamma} \frac{\pi T} {sin(\pi T\tau)},
\end{equation}
where $\epsilon_{\alpha}=(\epsilon_d+V_{\perp}k_D)$, $\Gamma=\rho t^2$ is the
resonance width of the $\alpha$ particle with the conduction electrons,
 and the phase shift
of $s$-electrons at the Fermi point due to the presence of the local impurity
$\alpha$ is found to be ${\frac{\pi}{2}- tan^{-1}(\frac{\epsilon_{\alpha}}
{\Gamma}})$. Generally, it is known that the impurity energy level is not
exactly pinned at the chemical potential, being very close to it. However,
for simplicity, we assume the extreme condition letting $\epsilon_{\alpha}=0$,
so the above phase shift is chosen $\pi/2$ approximately. In the subspace
$n_{\beta}=1$, the condition $n_{\alpha}=0$ is required
 by the local constraint, so
the  conduction electrons become free. More importantly, the essential
physics of this mixed valence quantum critical regime is contained in the Green
function of $\beta$ in the restricted Hilbert space, which can be expressed as
\begin{eqnarray}
&& G_{\beta}(\tau)= -<T_{\tau}\beta(\tau)\beta^{\dag}(0)> \nonumber \\ &&
\hspace{1cm}  = -\theta(\tau)
<\phi\mid\beta(\tau)\beta^{\dag}(0)\mid\phi>_{n_{\beta}=0}
 +\theta(-\tau)
 <\phi\mid\beta^{\dag}(0)\beta(\tau)\mid\phi>_{n_{\beta}=1}.
\end{eqnarray}
Due to the conservation of $n_{\beta}$, the first term can be  regarded as an
XRE  absorption process, while the second term as an  XRE emission process.
 From the well-known work on XRE problem \cite{nd}, the retarded Green's
function ${\beta}$ is proved to be
\begin{equation}
 G_{\beta}^R(\tau)\approx -\theta(\tau)
  [\frac{\pi T}{\Gamma sin(\pi T\tau)}]^{1/4}.
\end{equation}
This result shows that the single-particle Green function of $\beta$
impurity displays XRE singularities or Anderson orthogonality catastrophe
at this mixed valence quantum critical point. Based on these results, we find
that
\begin{equation}
 <n_{\beta}>\sim T^{1/4}, \hspace{1cm}
   <n_{\alpha}>= \frac{1}{2}, \hspace{1cm} T\rightarrow 0.
\end{equation}
Thus, we have $<n_d>=1/2$, and from the Friedel sum rule, the local impurity
{\it approximately } has a phase shift $\pi/4$ per spin at the Fermi point at
zero temperature. Moreover, the correlation function of the charge density
$\rho_d=n_{\alpha}+n_{\beta}-\frac{1}{2}$  can be evaluated as
\begin{equation}
<\rho_d(\tau)\rho_d(0)>\approx(\frac{1}{\pi\Gamma})^2
 [\frac{\pi/\beta}{sin(\pi\tau/\beta)}]^2,
  \hspace{.5cm} \tau>0.
\end{equation}
However, the correlation function of the spin density
$\sigma^z_d\equiv \alpha^{\dag}\beta+\beta^{\dag}\alpha$
is more subtle because of the local constraint. For $\tau>0$, it is written as
$$ <\sigma_d^z(\tau)\sigma_d^z(0)>=<\alpha^{\dag}(\tau)\beta(\tau)
\beta^{\dag}(0)\alpha(0)>_{n_{\beta=0}}+
 <\beta^{\dag}(\tau)\alpha(\tau)\alpha^{\dag}(0)\beta(0)>_{n_{\beta}=1}.$$
When the local constraint is carefully considered, the nonzero contribution to
the first expectation value is
$$<e^{H_0\tau}\alpha^{\dag}\beta e^{-H_1\tau}\beta^{\dag}\alpha>_{n_{\beta}=0}
 = <e^{H_0\tau}\alpha^{\dag}e^{-H_1\tau}\alpha>_{n_{\beta}=0},$$
while the nonzero contribution to the second term is
$$<e^{H_1\tau}\beta^{\dag}\alpha e^{-H_0\tau}\alpha^{\dag}\beta>_{n_{\beta}=1}
= < e^{H_1\tau}\alpha e^{-H_0\tau}\alpha^{\dag}>_{n_{\beta}=0}, $$
where $H_0$ is the effective Hamiltonian in the $n_{\beta}=0$ subspace,
and $H_1$ is the effective Hamiltonian in the $n_{\beta}=1$ subspace.
These expectation values can be evaluated in the same way as the retarded Green
function of $\beta$\cite{nd}. We thus derive the final result
\begin{equation}
<\sigma_d^z(\tau)\sigma_d^z(0)>\approx
2[\frac{1}{\Gamma}\frac{\pi T}{sin(\pi T\tau)}]^{1/4},
\hspace{.5cm} \tau>0.
\end{equation}
The longitudinal spin-spin correlation function is dramatically
different from the usual single-impurity model ($1/\tau^2$  in
the low temperature limit). Here we have assumed $\epsilon_d\approx 0$.
Since the local impurity decouples from the
magnetic field and hybridizing electrons at this limit, the singular
part of $\chi_{imp}$ and $C_{imp}$ vanishes exactly.
A perturbative calculation around the solvable strong coupling limit is
needed to explore the generic behavior of the mixed valence phenomena.

\subsection{Toulouse limit in the single-occupancy}

We now turn to the discussions away from the strong coupling limit. As a test,
we have to recover the well-known results of the single-impurity model in the
Kondo limit. Since $n_{d,\uparrow}+n_{d,\downarrow}=1$ in the Kondo lmit, we
have the following effective Hamiltonian:
\begin{eqnarray}
&& \tilde{H}=
\sum_{k>0,\sigma}\frac{k}{\rho}b^{\dag}_{k,\sigma}b_{k,\sigma}
+V_{\perp}k_D\sum_{\sigma}d^{\dag}_{\bar{\sigma}}d_{\sigma}
+(V_0-\frac{1}{\rho})\sum_{k>0,\sigma}\sqrt{\frac{k}{N}}
(b^{\dag}_{k,\sigma}+b_{k,\sigma}) (n_{d,\sigma}-\frac{1}{2})
\nonumber \\ && \hspace{1cm}
+{V'}_0\sum_{k>0,\sigma}\sqrt{\frac{k}{N}}(b^{\dag}_{k,\sigma}+b_{k,\sigma})
(n_{d,{\bar \sigma}}-\frac{1}{2})
\nonumber \\ && \hspace{1cm}
+ \frac{h}{4\pi}\sum_{k>0}\sqrt{\frac{k}{N}}\int dx
[(b_{k,\uparrow} - b_{k,\downarrow})e^{ikx}+
 (b^{\dag}_{k,\uparrow}-b^{\dag}_{k,\downarrow})e^{-ikx}],
\end{eqnarray}
where the hybridization term disappears due to the single occupancy constraint
and we have dropped the screening channel  because it decouples from the
the impurity. Moreover, the charge degrees of freedom for the hybridizing
electrons can also be separated out if we introduce the charge and spin-density
operators as $e_{k}=\frac{1}{\sqrt{2}}(b_{k,\uparrow}+b_{k,\downarrow})$
and $b_{k}=\frac{1}{\sqrt{2}}(b_{k,\uparrow}-b_{k,\downarrow})$, respectively.
After that, the effective Hamiltonian  for the Kondo phase is obtained as:
\begin{eqnarray}
&& H_{eff}^{K}=\sum_{k>0}\frac{k}{\rho}b^{\dag}_{k}b_{k}
+V_{\perp}k_D\sum_{\sigma}d^{\dag}_{\bar{\sigma}}d_{\sigma}
+\frac{1}{\sqrt{2}}(V_0-V'_0-\frac{1}{\rho})\sum_{k>0}\sqrt{\frac{k}{N}}
(b^{\dag}_{k}+b_{k})(n_{d,\uparrow}-n_{d,\downarrow})
\nonumber \\ && \hspace{1cm}
+ \frac{\sqrt{2}h}{4\pi}\sum_{k>0}\sqrt{\frac{k}{N}}
\int dx [(b_{k,m}e^{ikx}+ b_{k,m}^{\dag}e^{-ikx})].
\end{eqnarray}
It is very useful to perform the following canonical transformation
\begin{equation}
 P=exp\{- \frac{1}{2}\sum_{k>0}\frac{1}{\sqrt{kN}}
 (b_k-b_k^{\dag})(d^{\dag}_{\uparrow}d_{\uparrow}
 -d^{\dag}_{\downarrow}d_{\downarrow})\}.
\end{equation}
When we transform the bosons back to fermions, $H_{eff}^K$ becomes
\begin{eqnarray}
&& \tilde{H}_{eff}^{K}=\sum_{k}(\epsilon_k +
\frac{h}{\sqrt{2}})C^{\dag}_{k}C_{k} +V_{\perp}\sqrt{k_D}\sum_{k}
(C^{\dag}_kd^{\dag}_{\downarrow}d_{\uparrow} +
 d_{\uparrow}^{\dag}d_{\downarrow}C_k)
+\frac{\sqrt{2}h}{4}(n_{d,\uparrow}-n_{d,\downarrow})
\nonumber \\ && \hspace{0.5cm}
+[\frac{1}{\sqrt{2}}(V_0-V'_0-\frac{1}{\rho})+\frac{1}{2\rho}]
\frac{1}{2N}\sum_{k,k'}(C^{\dag}_{k}C_{k'}-C_{k'}C^{\dag}_{k})
(n_{d,\uparrow}-n_{d,\downarrow})
+\frac{\sqrt{2}h}{4}(n_{d,\uparrow}-n_{d,\downarrow}).
\end{eqnarray}
Note that due to the single-occupancy constraint
$n_{d,\uparrow}+n_{d,\downarrow}=1$, the local impurity is reduced to
$SU(2)$ spin-$1/2$ operators:
$$d^{\dag}_{\uparrow}d_{\downarrow}=S^+, \hspace{.5cm}
d^{\dag}_{\downarrow}d_{\uparrow}=S^-, \hspace{.5cm}
\frac{1}{2} (d^{\dag}_{\uparrow}d_{\uparrow}
 -d^{\dag}_{\downarrow}d_{\downarrow})=S^z.$$
In fact, the local impurity orbital can  also be expressed as a $U(1)$
spin-$1/2$ representations in terms of spinless fermions without
any constraints:
$$S^+=f^{\dag}, \hspace{.5cm} S^-=f, \hspace{.5cm}
S^z=f^{\dag}f-\frac{1}{2},$$
which is in fact the one-site Jordan-Wigner transformation for a spin-1/2
Pauli operator.

Thus, the final version of the Kondo effective Hamiltonian is
\begin{eqnarray}
&& \tilde{H}_{eff}^{K}=\sum_{k}(\epsilon_k +
\frac{h}{\sqrt{2}})C^{\dag}_{k}C_{k} +V_{\perp}\sqrt{k_D}\sum_{k}
(C^{\dag}_k f + f^{\dag}C_k)
+\frac{h}{\sqrt{2}}(f^{\dag}f -\frac{1}{2})
\nonumber \\ && \hspace{1cm}
+[\sqrt{2}(V_0-V'_0-\frac{1}{\rho})+\frac{1}{\rho}]
\frac{1}{2N}\sum_{k,k'}(C^{\dag}_{k}C_{k'}-C_{k'}C^{\dag}_{k})
(f^{\dag}f -\frac{1}{2}).
\end{eqnarray}
This is a single-channel resonant-level model \cite{wfs}, from which the
well-known
Toulouse limit Hamiltonian \cite{tou} can be deduced when
$V_0-V'_0=\frac{1}{\rho}( 1 - \frac{\sqrt{2}}{2} ) $:
\begin{equation}
 \tilde{H}_{eff}^{K}=\sum_{k}(\epsilon_k +
\frac{h}{\sqrt{2}})C^{\dag}_{k}C_{k} +V_{\perp}\sqrt{k_D}\sum_{k}
(C^{\dag}_k f + f^{\dag}C_k)+\frac{h}{\sqrt{2}}(f^{\dag}f -\frac{1}{2}).
\end{equation}
Here it is exactly derived from the Anderson single-impurity model
with the single-occupation constraint.
 Using the Toulouse limit conditions, the Fourier transformation of
the $f$-electron propagator $G_f(\omega)$ is evaluated as
\begin{equation}
G_f(\omega)=[\omega-\frac{h}{\sqrt{2}}+i\Gamma_K sign(\omega)]^{-1},
\end{equation}
where $\Gamma_K=\rho V^2_{\perp}k_D/2$.
The corresponding density of states is
$$ \rho_f (\omega)=\frac{1}{\pi} \frac{\Gamma_K}
{(\omega-\frac{h}{\sqrt{2}})^2+(\Gamma_K)^2}.$$
This expression brings out
the essential feature of the Kondo problem. In the strong coupling
regime, the spin degrees of freedom of the local impurity have been
quenched out by a conduction electron (forming a spin singlet), and
other conduction electrons just hybridize with the impurity charge
degrees of freedom with Lorentzian width $\Gamma_K$, or they only feel a
resonant scattering potential provided by the spin singlet at the
origin. Thus, the change of the free energy due to this hybridization may be
calculated by using the density of states:
\begin{equation}
\delta F_{imp}= \frac{1}{\pi}\int^{\infty}_{-\infty}d\omega
f(\omega)tan^{-1} (\frac{\Gamma_K}{\omega -\frac{h}{\sqrt{2}}}),
\end{equation}
where $f(\omega)=(e^{\beta\omega}+1)^{-1}$. Then, it is
straightforward to evaluate the impurity contribution to the entropy
$lim_{T\rightarrow 0}lim_{h\rightarrow 0}S_{imp}=0$,
the specific heat $C_{imp}=\frac{\pi}{3\Gamma_K}T$, and the spin
susceptibility $\chi_{imp}=\frac{1}{2\pi\Gamma_K}$. Thus, the known
universal Wilson ratio $R_W=
(\frac{T\chi_{imp}}{C_{imp}})/(\frac{T\chi_{bulk}}{C_{bulk}})=2$ is
recovered \cite{wilson,rn}.

\subsection{Generic properties in the mixed valence quantum critical
 regime}

Let us now concentrate on the physical properties of the mixed valence
regime around the  mixed valence quantum critical point.
The effective Hamiltonian is given by $H_{eff}^{MV}=H_0+\delta H$, where
\begin{eqnarray}
&& H_0=\sum_{k,\sigma} ( \epsilon_k + \frac{h}{2}\sigma)
 C^{\dag}_{k,\sigma}C_{k,\sigma}+\sum_{k}\epsilon_k s^{\dag}_{k}s_{k}+
 (\epsilon_d -V_{\perp}k_D) n_{\beta}
\nonumber \\ && \hspace{1cm}
+(\epsilon_{d}+ V_{\perp} k_D)n_{\alpha}
 +\sqrt{2}t (s_0^{\dag}\alpha^{\dag}+h.c.)
\nonumber \\ &&
\delta H=\frac{1}{2}(V_0+V'_0-\frac{1}{\rho})
\sum_{\sigma} (C^{\dag}_{\sigma}C_{\sigma}-\frac{1}{2})(\rho_d -\frac{1}{2})
+ \frac{1}{2}(V_0-V'_0-\frac{1}{\rho})\sum_{\sigma}\sigma
 C^{\dag}_{\sigma}C_{\sigma}\sigma^z_d
\nonumber \\ && \hspace{1cm}
+(\tilde{V}_s-\frac{1}{\rho}) (s^{\dag}_{0}s_{0}-\frac{1}{2})\rho_d,
\end{eqnarray}
where  $\epsilon_d\approx -V_{\perp}k_D$ and
$\delta H$ is the perturbation away from the mixed valence quantum critical
point. Obviously, there is a decoupling of charge and spin densities
for the hybridizing electrons which  makes our following calculations
transparent. After that separation, we get
\begin{eqnarray}
&& H_0=\sum_{k}(\epsilon_k + \frac{h}{\sqrt{2}})
 g^{\dag}_{k}g_{k}+\sum_{k}\epsilon_k p^{\dag}_{k}p_{k}+
\sum_{k}\epsilon_k s^{\dag}_{k}s_{k}+
 (\epsilon_{d}-V'_0k_D+V_{\perp}k_D)n_{\alpha}
\nonumber \\ && \hspace{1cm}
+(\epsilon_d -V'_0k_D-V_{\perp}k_D)n_{\beta}
+\sqrt{2}t (s_0^{\dag}\alpha^{\dag}+h.c.)
\nonumber \\ &&
\delta H=\frac{1}{\sqrt{2}}(V_0+V'_0-\frac{1}{\rho})
 (p^{\dag}_0p_0-\frac{1}{2})(\rho_d-\frac{1}{2})+
\frac{1}{\sqrt{2}}(V_0-V'_0-\frac{1}{\rho})
 (g^{\dag}_0g_0-\frac{1}{2})\sigma^z_d
\nonumber \\ && \hspace{1cm}
+(\tilde{V}_s-\frac{1}{\rho})(s^{\dag}_0s_0-\frac{1}{2})\rho_d.
\end{eqnarray}
Here the spinless $p-$ and $g-$electrons describe the charge and spin
degrees of freedom of the hybridizing electrons, respectively. $H_0$
corresponds to our mixed valence quantum critical point in the strong
coupling limit. As follows from our calculated
charge density and spin
density correlation functions of the local impurity of $H_0$, only
interactions associated with the spin density correlation function in the
perturbations are singular. We also know the local propagator of the spin
part of the hybridizing electrons
$$ G_g(\tau)= -<T_{\tau}g^{\dag}_0(\tau)g_0(0)>
    =k_D T\sum_n\int dk \frac{e^{i\omega_n\tau}}
              {i\omega_n-\frac{k}{\rho}-\frac{h}{\sqrt{2}}}
    \approx\rho \frac{\pi T}{sin(\pi T\tau)},$$
but $G_g(0)\approx \frac{\rho h}{\sqrt{2}}$ for $h\rightarrow 0$.

To the second order in perturbation, the singular contribution of
$\delta H$ to the impurity free energy reads as:
\begin{eqnarray}
&& \Delta F_{imp}= -
\frac{1}{2}\int_0^{1/T}d\tau <\delta H(\tau)\delta H(0)>
\nonumber \\ && \hspace{1cm}
=-\frac{\lambda^2}{2} \{
 G^2_g(0)\int_0^{1/T}d\tau<\sigma^z_d(\tau)\sigma^z_d(0)> +
 \int_0^{1/T}d\tau G^2_g(\tau) <\sigma^z_d(\tau)\sigma^z_d(0)> \},
\end{eqnarray}
where $\lambda=\frac{1}{\sqrt{2}}(V_0-V'_0-\frac{1}{\rho})$.
It turns out that the first term yields the dominant contribution to
$\chi_{imp}$, while the second term yields the dominant contribution to
$C_{imp}$. At low temperatures ($T\ll\Gamma$), the asymptotic forms of the
propagators can be used in the expression of $\Delta F_{imp}$ to yield
\begin{equation}
 \Delta F_{imp}=-\frac{2\lambda^2\rho^2h^2}{\Gamma}
 (\frac{\Gamma}{\pi T})^{3/4} \int_{0}^{1/2T}
 \frac{\pi T}{sin^{1/4}(\pi T\tau)} d\tau
 -2\lambda^2 \rho^2 \Gamma (\frac{\pi T}{\Gamma})^{5/4}\int_{0}^{1/2T}
 \frac{\pi T}{sin^{9/4}(\pi T\tau)} d\tau.
\end{equation}
All integrals must be regularized by introducing an ultraviolet cut-off
$\Gamma^{-1}$. Finally, the
low-temperature singular behavior of the spin susceptibility and specific heat
is found to be
$$\chi_{imp}=
  \frac{2\lambda^2\rho^2}{\Gamma}A(\frac{\Gamma}{\pi T})^{3/4}, \hspace{.5cm}
  C_{imp}=\frac{\pi\lambda^2\rho^2}{16}A(\frac{\pi T}{\Gamma})^{1/4},$$
where $A\equiv\frac{\Gamma(3/8)\Gamma(1/2)}{\Gamma(7/8)}$ and
$\Gamma(x)$ is the  Euler Gamma function. Higher order perturbation
terms in $\lambda$ yield subdominant contributions at low temperatures, and
will not modify these results. Therefore, the behavior of susceptibility and
specific heat obtained here is generic in the mixed valence regime. The spin
susceptibility of the local impurity explicitly reveals the mixed valence
physics around the strong coupling limit: {\it not fully quenched magnetic
moment and exhibition of the XRE singularities}. Put it  another way,
{\it the local moment fluctuations are essential for the mixed
valence state}. When we consider the leading power-law singularity of the
spin susceptibility and specific heat, we find the effective Wilson ratio
$T\chi_{imp}/C_{imp}=32/{\pi^2}$, which should be observable in experiment.
The present method allows us to see explicitly how the dominant singular
behavior of the thermodynamic quantities is governed by the leading irrelevant
interactions in the strong coupling  Hamiltonian, and thereby correctly
captures the generic properties of the generalized Anderson single-impurity
model in the mixed valence phase.

\section{Experiments on $UP\lowercase{d}_{\lowercase{x}}C
\lowercase{u}_{5-\lowercase{x}}$ alloys}

Now we try to relate the  above obtained results  to real strongly
correlated f-electron heavy fermion systems, {\it e.g.} $UPd_xCu_{5-x}$ alloys
in which U ions have partially-filled f-electron shells and give rise
to magnetic
moments interacting with the spin and charge of the conduction electrons.
What is striking about these materials is the non-FL behavior of their
physical properties which exhibit weak power law or logarithmic divergences in
temperature and suggest the existence of a critical regime at $T=0$
\cite{rev,andraka,aronson,tsve}.

The parent
compound $UCu_5$ is a prototype Kondo impurity system at high temperatures.
As temperature is lowered, the intersite interactions drive a transition to a
long-range antiferromagnetic (AFM) order of incompletely compensated uranum
moments at $15K$. But as Pd is substituted on the Cu sites, the AFM order
is suppressed, with the ordering temperature being driven to zero.
 In particular,
the recent neutron scattering experiments \cite{aronson} showed
that the magnetic
response function for x=1 and x=1.5 has no appreciable momentum
(except for the ionic form factor) and
temperature dependence and has a scale invariant form
$S(\omega)\sim \omega^{-1/3}$ at  very low energy scale (less than 10 mev).
 Moreover, the specific
heat and spin susceptibility measurements \cite{andraka} also showed that
between 1K and 10K the temperature variations for x=1 and x=1.5
$C_{imp}/T \sim \chi_{imp}$ apparently deviate from the logarithmic temperature
dependence (expected for  a two-channel Kondo model at low
temperatures), and can be much better represented by a power law
$C_{imp}/T \sim \chi_{imp} \sim T^{-\Delta}$ with $\Delta=0.32$. Furthermore,
the neutron data
at various energies and temperatures\cite{aronson} have been fitted
by a single scaling function, and it has been argued that the existence
of such a scaling behavior provides a strong evidence in favor of
single-impurity critical phenomena\cite{tsve}. The conformal dimension
of the spin correlation extracted from this fitting $\Delta = 1/3$ is
consistent with  the specific heat, magnetic susceptibility and resistivity
data\cite{tsve}. On the other hand, it is very clear that neither the
single channel Kondo (predicting the FL behavior) nor the two-channel Kondo
(leading to logarithmic singularities) can provide any acceptable
explanation for
such a striking behavior. It is thus natural to expect that the additional
charge fluctuations inherent in the mixed valence phenomena, may shed
some light on this puzzle.

Here we propose a possible explanation for these experimental observations
in terms of the generalized Anderson model considered in this paper.
First, $UPd_xCu_{5-x}$ with x=1 and x=1.5 may be described as
 mixed valence states because the f-electrons of U-ion is only partially
filled and the impurity energy level is split
in the crystalline field into a large
number of sublevels  with possibly one of them
being  close to the chemical potential of the
conduction electrons. Second, there are probably strong screening interactions
from the off-site conduction electrons and strong impurity-phonon interaction
due to the large lattice distortion accompanying the impurity occupancy
changes. Finally, due to the presence of strong screening and
phonon interactions, the motion of the U-ion must drag a heavy screening
cloud with it.
Hence its motion is  very "slow", so the correlations from the inter-impurity
interaction are rather weak. Thus, we can use our generalized Anderson
single-impurity model to describe these materials.

According to our  calculations, we have found that the
extra specific heat and spin susceptibility in the mixed valence quantum
critical regime  can be expressed as
$C_{imp}/T \sim \chi_{imp} \sim T^{-3/4}$ in the strong-coupling limit.
As for the magnetic response function $S(\omega)$ from the neutral scattering
measurements, we can obtain it after performing Fourier transformation for the
spin-spin density correlation derived in the strong coupling limit of the
mixed valence regime. We found that in the extremely low-energy scale
$S(\omega)\sim \omega^{-3/4}$. Apparently, the singularity
exponent derived from our theoretical calculations (-3/4) is stronger than
the experimental results $\sim -1/3$. A possible
way to recoincide these results is to note that
in our treatment we have pushed the impurity scattering to the
unitarity limit (phase shift  $\pi/2$), while in realistic
systems the unitarity limit may not be reached, so the
exponents might be different\cite{tsve1}. In view of the experimental
relevance, it is very important
to find other ways to treat this  intriguing problem for cross-checking.

\section{Concluding remarks}

To summarize, in this paper we have considered a generalized Anderson model
with screening channels, which can be considered as a single-site version
of the three-band Hubbard model originally proposed to describe the
cuprates. Using the bosonization technique and canonical
transformations we have found
the strong coupling Toulouse limit of the effective Hamiltonian which
can be solved exactly. The physical properties around the mixed valence
quantum critical point separating the two FL phases (the Kondo and the
empty orbital phases) have been calculated explicitly
 to predict a new universality
class of non-FL behavior.

Our strong coupling treatment of the mixed valence problem
provides the following physical picture for the mixed valence state. In the
hybridizing channel, the phase shift due to hybridization is compensated
by the parallel-spin XRE scattering via transformation $S$, while the XRE
scattering in the screening channel is converted into an effective
hybridization with the local impurity via transformation $U$. However,
only "half" of the local impurity ($\alpha$ particle) hybridizes with the
screening electrons (in fact "holes") and gives rise to a standard FL behavior.
The other "half" of the local impurity ($\beta$ particle) does not mix with
the screening electrons and its expectation value $<n_{\beta}>\sim T^{1/4}$
at low temperatures. Nevertheless, its dynamic fluctuations lead to XRE
singularities and a divergent power-law susceptibility which are
consistent with observations
in $UPd_xCu_{5-x}$ alloys with x=1 and x=1.5.

The non-trivial physics
shown in this paper is grasped due to a careful treatment of the
single occupancy constraint in the Toulouse limit. The essence of this
constraint is to satisfy the unitarity condition, i.e., to use a complete
set in the Hilbert space. Any mean-field type treatment of the constraint
will certainly miss this basic point.

 We would like to thank M. Aronson, M. Fabrizio, T. Giamarchi,  A. C. Hewson,
 G. Kotliar, Y. L. Liu, and  A.M. Tsvelik for helpful discussions.
 G.-M. Zhang is supported by the Science
and Engineering Research Coucil (SERC) in Britain.

\end{document}